\documentclass[openacc]{rstransa}%%%%where rstrans is the template name

\usepackage{bbold}
\usepackage{subcaption}
\usepackage{bm}
\begin{document}

\newcommand{\phil}[1]{\textcolor{red}{#1}}
\newcommand{\philc}[1]{\textcolor{red}{[#1]}}
\newcommand{\tsh}[1]{\textcolor{blue}{#1}}
\newcommand{\tshc}[1]{\textcolor{blue}{#1}}
\newcommand{\gb}[1]{\textcolor{red}{#1}}
\newcommand{\gbc}[1]{\textcolor{red}{[#1]}}

%%%% Article title to be placed here
\title{Simulations of Frustrated Ising Hamiltonians using Quantum Approximate Optimization}

\author{%%%% Author details
Phillip~C.~Lotshaw$^{1}$, Hanjing Xu$^{2}$, Bilal Khalid$^{2,3,4}$, Gilles Buchs$^{1,4}$, Travis~S.~Humble$^{1,4}$ and Arnab Banerjee$^{2,3,4}$
}

%%%%%%%%% Insert author address here
\address{$^{1}$Quantum Information Sciences Section, Oak Ridge National Laboratory, Oak Ridge, TN 37831, USA\\
$^{2}$ Purdue Quantum Science and Engineering Institute, Purdue University, West Lafayette, IN 47907, USA
\\
$^{3}$ Department of Physics and Astronomy, Purdue University, West Lafayette, IN 47907, USA
\\
$^{4}$ Quantum Science Center, Oak Ridge National Laboratory, Oak Ridge, TN 37830, USA}

%%%% Subject entries to be placed here %%%%
\subject{Quantum Computing}

%%%% Keyword entries to be placed here %%%%
\keywords{quantum simulation}

%%%% Insert corresponding author and its email address}
\corres{Phillip C. Lotshaw\\
\email{lotshawpc@ornl.gov}\\
------------------

This manuscript has been authored by UT-Battelle, LLC under Contract No. DE-AC05-00OR22725 with the U.S. Department of Energy. The United States Government retains and the publisher, by accepting the article for publication, acknowledges that the United States Government retains a non-exclusive, paid-up, irrevocable, world-wide license to publish or reproduce the published form of this manuscript, or allow others to do so, for United States Government purposes. The Department of Energy will provide public access to these results of federally sponsored research in accordance with the DOE Public Access Plan. (http://energy.gov/downloads/doe-public-access-plan)}

%%%% Abstract text to be placed here %%%%%%%%%%%%
\begin{abstract}
Novel magnetic materials are important for future technological advances.  Theoretical and numerical calculations of ground state properties are essential in understanding these materials, however, computational complexity limits conventional methods for studying these states. Here we investigate an alternative approach to preparing materials ground states using the quantum approximate optimization algorithm (QAOA) on near-term quantum computers.  We study classical Ising spin models on unit cells of square, Shastry-Sutherland, and triangular lattices, with varying field amplitudes and couplings in the material Hamiltonian. We find relationships between the theoretical QAOA success probability and the structure of the ground state, indicating that only a modest number of measurements ($\lesssim100$) are needed to find the ground state of our nine-spin Hamiltonians, even for parameters leading to frustrated magnetism. We further demonstrate the approach in calculations on a trapped-ion quantum computer and succeed in recovering each ground state of the Shastry-Sutherland unit cell with probabilities close to ideal theoretical values. The results demonstrate the viability of QAOA for materials ground state preparation in the frustrated Ising limit, giving important first steps towards larger sizes and more complex Hamiltonians where quantum computational advantage may prove essential in developing a systematic understanding of novel materials. 
\end{abstract}
%%%%%%%%%%%%%%%%%%%%%%%%%%%

\maketitle

\section{Introduction}
Quantum magnetism has been a major focus in condensed matter research, driven by the potential for new disruptive applications ranging from quantum computing to quantum sensing \cite{blundell_2014}. Quantum material properties are intrinsically related to the structure of the ground states. However, exact ground states are notoriously challenging to calculate classically, requiring the field to resort to using semi-classical limits \cite{samarakoon_banerjee_zhang_kamiya_nagler_tennant_lee_batista_2017, 
winter_li_jeschke_valenti_2016, Vojta_2018,  broholm_cava_kivelson_nocera_norman_senthil_2020} or fully quantum approaches with restricted applicability \cite{zhu_chen_zhang_li_jiang_wu_zhang_2019, szasz_motruk_zaletel_moore_2020,wessler_roessli_kramer_delley_waldmann_keller_cheptiakov_braun_kenzelmann_2020,
stoudenmire_white_2012, Wessel_2017, kamiya_kato_nasu_motome_2015, Troyer_2005}.  New, fully quantum computational tools are required to understand current problems including frustrated two-dimensional quantum magnets currently explored by bulk neutron scattering and thin film susceptibility \cite{Samarakoon_2022, Zhu_2017 }.  Digital and analog quantum simulators have emerged as a new tool for the simulation of quantum many-body phenomena towards efficient modeling of exotic quantum phases of matter beyond classical tractability \cite{Altman_2021_PRXQuantum,Georgescu_2014_RMP}.  They are naturally suited for magnetic Hamiltonians since spins can be directly mapped to qubits.  Non-trivial phases in magnetic systems, such as frustrated phases \cite{king_nisoli_dahl_poulin-lamarre_lopez-bezanilla_2021} , spin glasses \cite{harris_sato_berkley_reis_altomare_amin_boothby_bunyk_deng_enderud_etal_2018}, and topologically ordered phases \cite{bluvstein_levine_semeghini_wang_ebadi_kalinowski_keesling_maskara_pichler_greiner_etal_2022,satzinger_liu_smith_knapp_newman_jones_chen_quintana_mi_dunsworth_etal._2021} have been realized on multiple qubit platforms using a variety of techniques.
\par
In this paper, we investigate an alternative approach to preparing materials ground states using the quantum approximate optimization algorithm (QAOA) \cite{farhi2014quantum} on near-term quantum computers.  We apply QAOA to lattices of interest in materials science, considering the classical Ising limit (equivalently, $S=\infty$) where standard QAOA is directly applicable.  This serves as a stepping stone towards truly quantum problems such as the $XY$ and Heisenberg models in the fully frustrated limit, which will require further algorithmic research and modifications to the approach presented here.  Our results validate that QAOA achieves sufficient accuracy for the simpler classical limit and provides insights into algorithmic behavior for material lattice problems.  
\par
\begin{figure*}
\centering
\includegraphics[width=3cm,height=10cm,keepaspectratio]{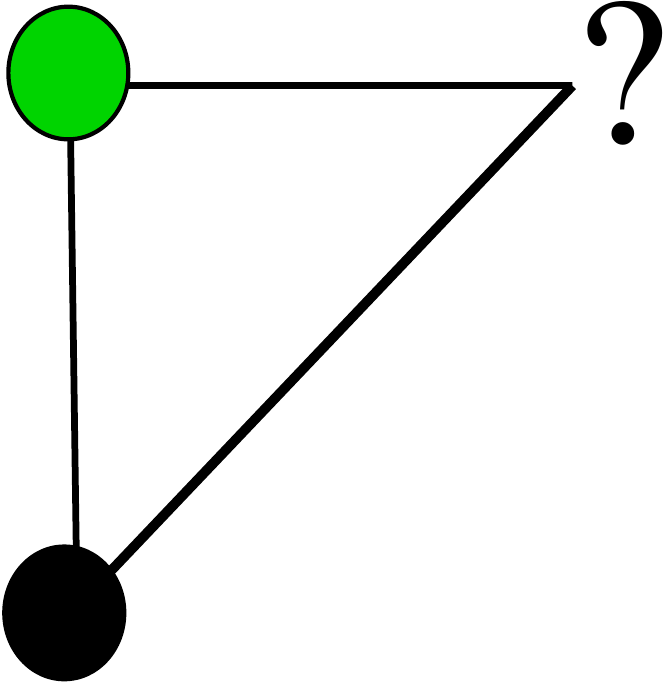}
\caption{Example of frustration on an anti-ferromagnetic triangular motif.  Two spins in opposite orientations (black and green) minimize the energy along one bond, however, there is no configuration for the final spin that minimizes energy along both remaining bonds. }
\label{frustration}
\end{figure*}

We consider lattice instances with varying degrees of frustration.  The smallest building block of a frustrated magnetic Hamiltonian is an anti-ferromagnetic triangular motif of three spins where all the bonds cannot be satisfied simultaneously. In these materials, exchange interactions compete such that it is impossible to satisfy them all simultaneously, see Fig.~\ref{frustration}. If all spin configurations are equally favorable, frustration can lead to non-ordered states such as spin liquids \cite{broholm_cava_kivelson_nocera_norman_senthil_2020}, spin glasses \cite{Ortiz-Ambriz_2019}, or plaquette states \cite{wessler_roessli_kramer_delley_waldmann_keller_cheptiakov_braun_kenzelmann_2020}, each with distinct signatures.
\par
\begin{figure*}
\centering
\includegraphics[width=\textwidth,height=10cm,keepaspectratio]{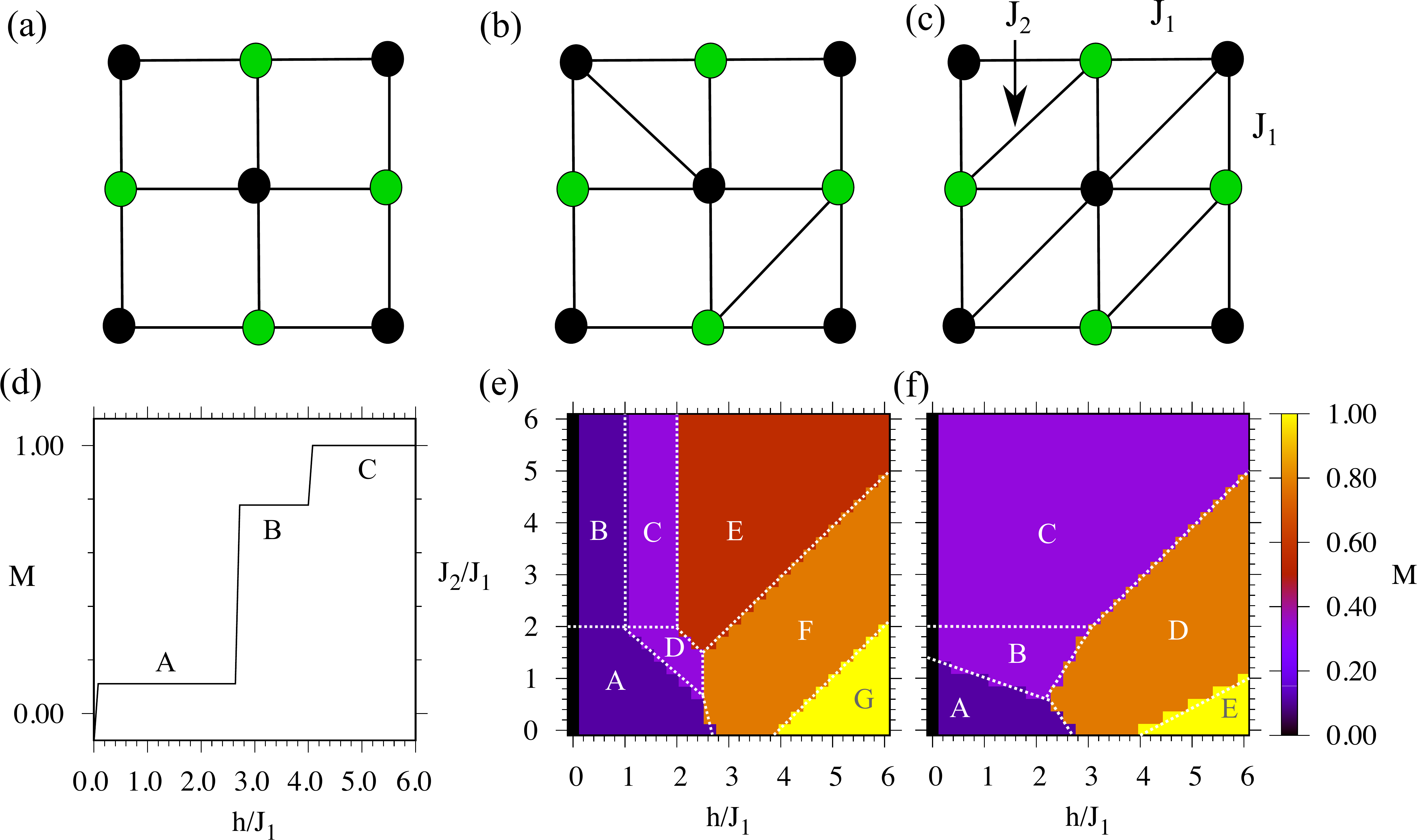}
\caption{Unit cells of (a) square, (b) Shastry-Sutherland, and (c) triangular geometries. Colors indicate two spin values $s_i = \pm 1$ in examples of ground states with $h/J_1,J_2/J_1, \ll 1$. (d-f) phase diagrams for each of the unit cells (a-c) respectively, with labels "A", "B",... denoting regions with distinct ground states for each lattice. Magnetization $M=0$ at $h=0$ is due to degeneracy in the ground states, where spin-flip-related pairs of states are present in absence of the field ($h=0)$.}
\label{lattices}
\end{figure*}

We solve three different types of Hamiltonians for unit cells pictured in Fig.~\ref{lattices}. The first is a square unit cell Hamiltonian, which exhibits only simple ferromagnetic and anti-ferromagnetic phases in the infinite size (thermodynamic) limit. The second is the celebrated Shastry-Sutherland lattice. Interestingly, this problem already lends itself to materials applications and experimental data analysis. Among other examples, it is conjectured to describe the class of rare-earth tetraborides  (ErB$_4$, TmB$_4$ and NdB$_4$) and allows a direct comparison with several existing results both theoretical \cite{ Dublenych_2012, Huang_2012, Kairys_2020, jha_stoyanoff_khundzakishvili_kairys_ushijima-mwesigwa_banerjee_2021, Farka_2010} and experimental \cite{Trinh_2018,Siemensmeyer_2008,ye_2017,panfilov_grechnev_zhuravleva_fedorchenko_muratov_2015, kim_sung_kang_kim_cho_rhyee_2010,Yoshii_2009}. The third case is the more complex Ising triangular lattice which represents a maximally frustrated problem with an infinite number of possible ground states in the infinite size limit \cite{shokef_souslov_lubensky_2011, Dublenych_2013 }. We compute theoretical probabilities to prepare the ground state for each of these 9-spin Hamiltonians under varying choices of the external field and coupling parameters and compare these theoretical results against computations on a trapped ion quantum computer. 
\par
We choose $N$ = 9 spins as this is the logical minimum number of spins required to construct a unit cell of the Shastry Sutherland lattice and also it is a feasible size for the Quantinuum quantum computer (with 12 qubits available at the time this work was completed). We focus on $p=1$ layers of the QAOA algorithm; for larger $N$ instances more layers $p$ of QAOA will be needed to maintain a significant success probability \cite{Lotshaw2021empirical,biamonte2022depthscaling,Ho2019variationalstateprep}.  Implementations on quantum computers will also have to overcome predicted limitations due to noise \cite{Coles2020nibp,Weidenfeller2022scaling,GonzalezGarcia2022propogation,GarciaPatron2021limitations} including an exponential scaling in the number of measurements with circuit size, depth, and other factors \cite{Lotshaw2022scaling}. Noise in modern quantum computers has negative consequences for all quantum algorithms, not only QAOA. Ongoing testing and development of these devices is necessary to assess realistic performance scaling in the presence of noise and to determine whether QAOA, or other quantum algorithms, will ultimately succeed in providing a useful computational advantage over conventional approaches. 
\section{Ising Hamiltonian and Model Unit Cell Lattices}
A single unpaired spin on the outermost orbital of a magnetic ion constitutes a $s$ = 1/2 state which is implemented straightforwardly on a qubit. In a magnetic material, several such spins in a lattice interact via pairwise superexchange interactions $J_\alpha$. The nature and strength of these interactions, $J_\alpha$, depend on several factors. These include the distance between the magnetic ions (typically $J_\alpha$ scales as the inverse cube of the distance between the ions), the shape of the orbitals, the symmetry of the lattice, and the local crystal fields. Magnetic frustration can arise in the lattice, for example, if spins arranged on a triangular motif in the lattice experience equal Ising anti-ferromagnetic interactions. Such a magnetic frustration can arise via a combination of straight edge bonds $J_1$ which are either horizontal or vertical, and the diagonal bonds $J_2$. This is given by the Hamiltonian
\begin{equation}\label{General Ising Model}
    \mathcal{H}(\bm s) = J_1\sum_{(i,j) \in \mathrm{NN}} s_{i}s_{j} + J_2\sum_{(i,j) \in \mathrm{NNN}} s_{i}s_{j} + h\sum_{i=1}^N s_i,
\end{equation}
 where the first sum is over the nearest neighbors (NN), the second sum is over the diagonal next-nearest neighbors (NNN), and $\bm s = (s_1, ..., s_N)$ lists the spin orientations $s_i \in \{1,-1\}$ of the $N$ spins.  We study anti-ferromagnetic couplings with positive $J_1$, $J_2$. The term $h$ represents a longitudinal magnetic field (parallel to the spin axis), which for the real material represents either a mean crystal field or an external magnetic field. The unit cell motif of the Hamiltonian is shown in Fig. \ref{lattices}. Materials described by this model are being actively researched in condensed matter physics. The Ising Shastry-Sutherland model is a special case of a model, inspired by the geometry of real materials, where some but not all of the diagonal bonds are present (Fig.~\ref{lattices}(b)). The triangular lattice is shown in Fig.~\ref{lattices}(c). In all cases we consider open boundary conditions. Analytical ground state properties of Ising models on Shastry-Sutherland and triangular lattices have been derived analytically  \cite{Shastry1981SSmodel,Brandt1986triangularising}.
\par
Multiple methods have been proposed to solve for ground states of Ising Hamiltonians and related optimization problems, notably Integer Programming method \cite{billionnet_elloumi_2006}, Simulated Annealing \cite{kirkpatrick_gelatt_vecchi_1983} and its variants, Large Neighborhood Search \cite{selby_2014} and Quantum or Quantum-inspired physical annealing devices. Among them, the Integer Programming method solves exactly but suffers from exponential scaling of computational time.  Simulated Annealing and Large Neighborhood Search are both herustics methods that promise faster runtime but there is no guarantee of the solution qualities. Quantum annealers, digital annealers and coherent Ising machines are hardwares dedicated to solving Ising models \cite{king_carrasquilla_raymond_ozfidan_andriyash_berkley_reis_lanting_harris_altomare_etal_2018, semeghini_levine_keesling_ebadi_wang_bluvstein_verresen_pichler_kalinowski_samajdar_etal_2021}. Depending on the connectivity of these various (qu)bits, every backend could be good at a different task - parallel tempering machines could be good in finding the classical phases \cite{jha_stoyanoff_khundzakishvili_kairys_ushijima-mwesigwa_banerjee_2021}, while others could reveal intricate dynamical behaviour in a transverse field Ising universality \cite{PhysRevResearch.2.033369}. For frustrated lattice problems, QAOA allows us to sample different ground states with certain probabilities due to quantum randomness, whereas classical and deterministic algorithms may generate a ground state efficiently, but fail to explore the states which might arise because of a coherent superposition between all the spins. Additionally, future extensions beyond the Ising limit also becomes an exciting possibility.

\subsection{Ground state magnetization phase diagrams}\label{phase diagrams}
We consider the nine-spin unit cells with geometries in Fig.~\ref{lattices}(a-c) which represent the number of spins required to simply construct an unit cell of the Shastry-Sutherland lattice. In materials represented by Bravais lattices, these unit cells repeat periodically to realize the very large lattices in a real material. We computed ground states for each unit cell by evaluating (\ref{General Ising Model}) for each possible spin configuration to identify the lowest energy states, for varying choices of $h$ and $J_2$, with $J_1=1$ taken as the unit of energy.  We plot the magnetization 
\begin{equation} M = \frac{1}{N} \sum_{i=1}^{N} s_i ,\end{equation}
of these ground states in the phase diagrams of Fig.~\ref{lattices}(d-f). We further separate each diagram into regions A,B,... with distinct sets of ground states but sometimes equivalent magnetizations.  For example, "A" and "B" in Fig.~\ref{lattices}(e) have different ground states but identical magnetizations. The individual ground states are shown in the Supplemental Information (Figs.~S4-S6) \cite{SI}.
\par
Starting with the non-frustrated square lattice with $J_1$ - only interactions with simple ferromagnetic or antiferromagnetic ground states, the degree of frustration is tuned progressively by a) addition of $J_2$ bonds and b) bringing $J_2 \to J_1$. The triangular lattice with uniform coupling parameters represents the maximally frustrated limit with highly degenerate solutions. The Shastry-Sutherland model represents a scenario with the minimum number of $J_2$ bonds required to realize a fully frustrated lattice. The solution of these states represent a problem of polynomial time complexity in two-dimensions and without a magnetic field. 
\par
The ground state for a given $h$ shows a number of magnetization plateaus, where each plateau has a different proportion of spins pointing up. Unsurprisingly, at large $h$, the ground state for each lattice is ferromagnetic, in regions C, G, and E in Fig.~\ref{lattices}(d-f) respectively. The situation becomes more interesting at small $h$ in regions "A", the ground state is anti-ferromagnetic with magnetization $M = 1/9$, as five spins are aligned with the field while four are anti-aligned.  For fields $8/3\leq h \leq 4$ and small $J_2$, there is a ground state with $M=7/9$ in which a single spin in the center of the unit cell is anti-aligned with the field. Besides frustration, these states are also determined by the finite size of the unit cells, where the central spin is distinguished as the only spin with four interactions in the square lattice. As $J_2$ and $h$ are varied, frustration leads to a variety of different ground states for the Shastry-Sutherland and triangular lattices, with varying magnetizations in Fig.~\ref{lattices}(e-f), with ground states in Supplemental Information \cite{SI}. These are true ground states of the 9-spin Hamiltonians, with boundary spins playing a big role. In the infinite size limit we expect the ground states to be progressively less dependent on the boundary, and more on the symmetry of the Hamiltonian, which we discuss in the next section. 

\subsection{Finite size effects} \label{finite size}
\begin{figure*}
    \centering
    \includegraphics[width=10cm,height=10cm,keepaspectratio]{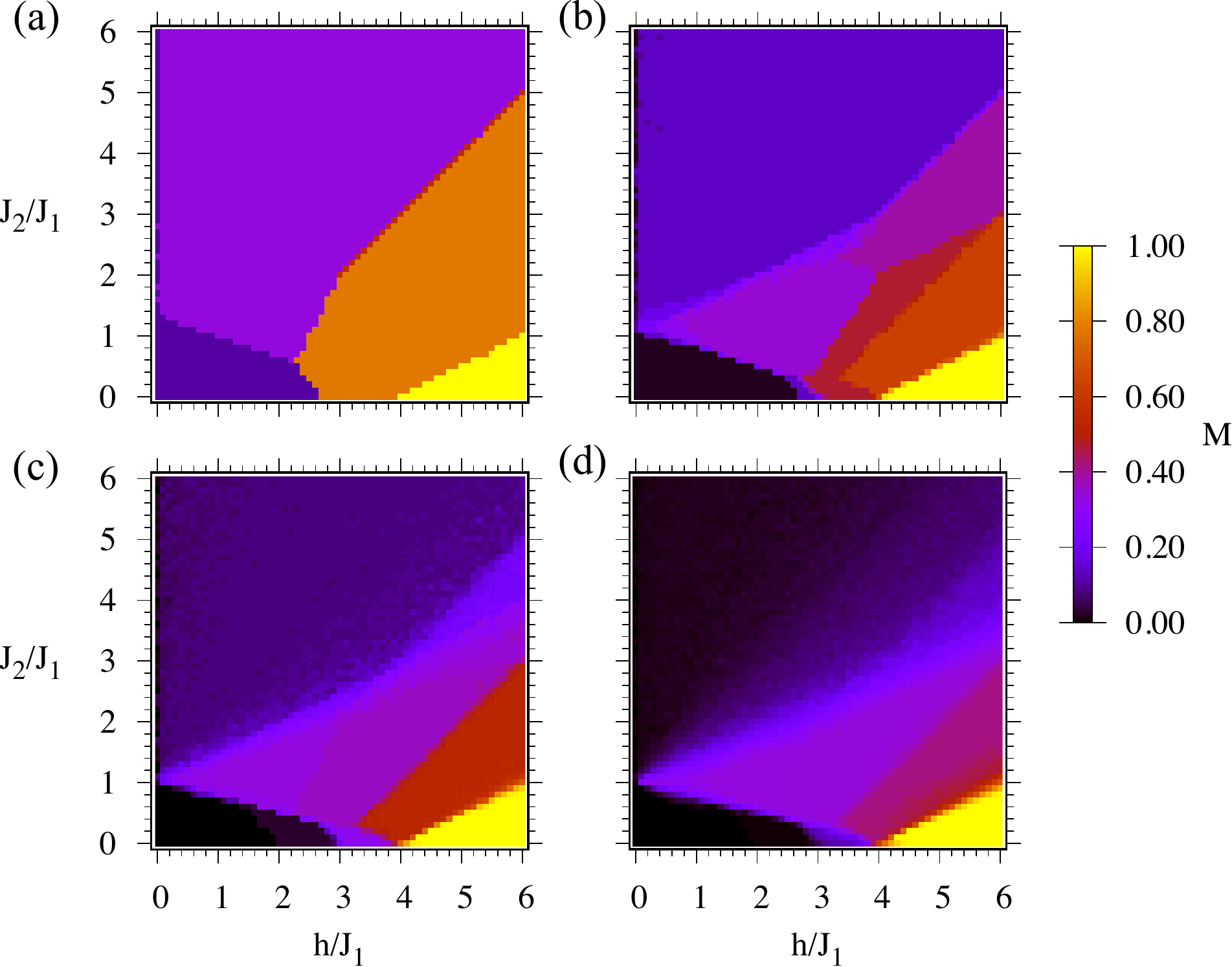}
\caption{Ground state magnetization of $n\times n$ triangular spin arrays with a number of spins per dimension (a) $n=3$, (b) $n=7$, (c) $n=12$, and (d) $n=30$, computed as described in Sec.~2\ref{finite size}. }
\label{Phase diagram size examples}
\end{figure*}

\begin{figure*}
\centering
    \includegraphics[width=8.5cm,height=10cm,keepaspectratio,trim={0cm 0 0 0},clip]{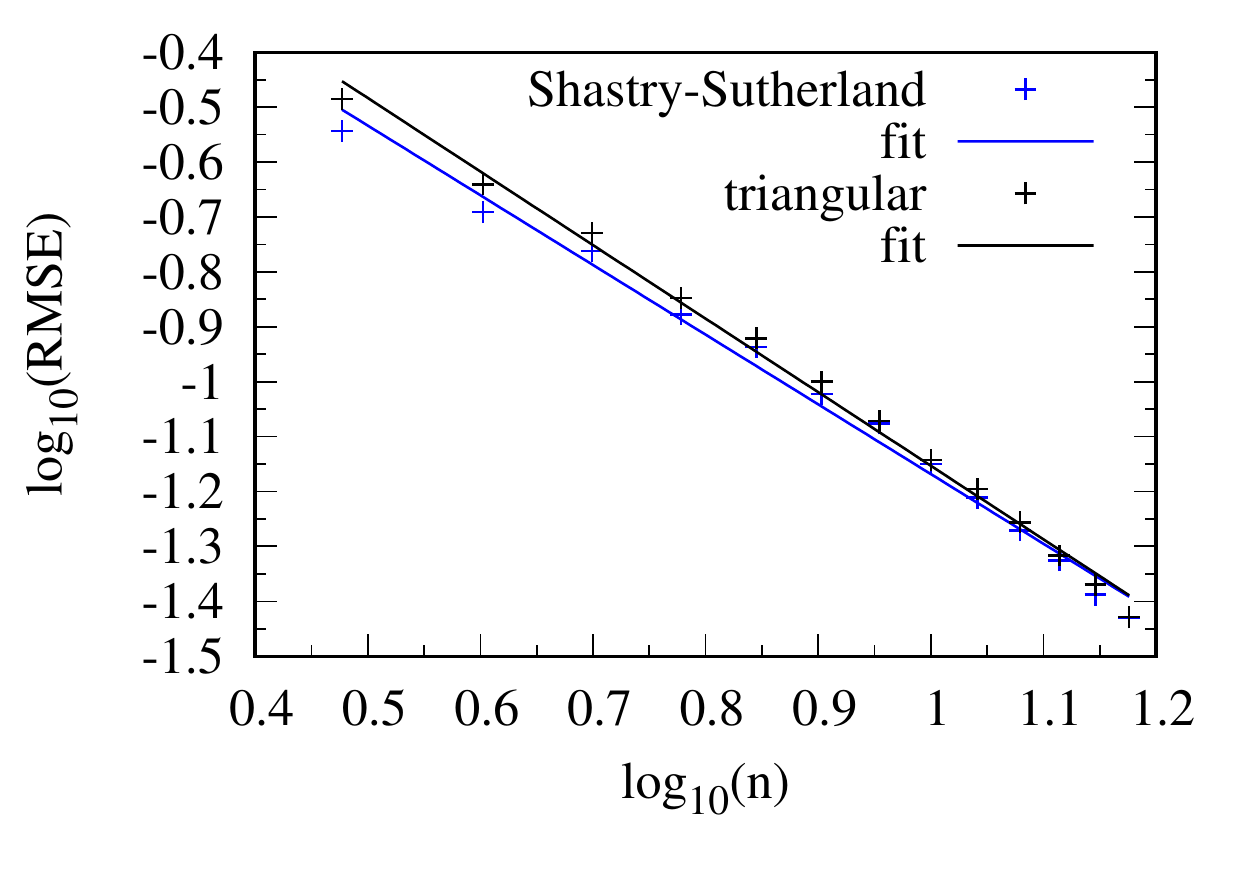}
\caption{ Root mean square errors (\ref{RMSE}) of total magnetization between  $n\times n$ lattices relative to a $30\times 30$ lattice, plotted on a log-log scale. Solid lines show fits to the power-law scaling relation (\ref{RMSE}); the slopes indicate the best-fit exponents $\gamma = -1.34(4)$ and $\gamma = -1.27(4)$ , with fit $R^2 =$ 0.994 and 0.990 for the triangular and Shastry-Sutherland lattices, respectively.  Best-fit intercepts are $\gamma\log_{10}(a)= 0.19$ and $\gamma\log_{10}(a)=0.10$ for the triangular and Shastry-Sutherland lattices respectively. The fit and the errors in the exponent are based on standard Levenberg-Marquardt routines and assume Poisson statistics at each point.}
\label{tri_finite_size_exponential_scaling}
\end{figure*}

The finite sizes of our lattice unit cells, as well as the unusual $M=7/9$ ground state noted in the previous section, raise questions of how the ground states for our unit cells match with ground states that would be obtained in the large size limit, and the minimum number of spins that are needed to achieve quantitative behavior consistent with large sizes.  To address these questions, we computed the magnetization of triangular and Shastry-Sutherland  lattices of $N = n\times n$ spins to analyze the size-dependent behavior. Due to the exponential complexity of the problem, we used \textit{neal} \cite{Dwave_neal}, a software implementation of simulated annealing to approximate the ground state configurations with $h, J_2 \in [0,6]$ and $J_1=1$. Each combination of $h$ and $J_2$ was run 50 times and the solution with minimum Ising energy was picked. Examples with $3 \leq n \leq 30$ are shown in Fig.~\ref{Phase diagram size examples}. We assessed convergence of the global phase diagrams to the large size limit by computing the root mean square error (RMSE) between the target lattice's and 30$\times$30 lattice's magnetization across points in the phase diagram. We fitted the RMSE to both a power-law as well as an exponential form (see Supplemental Information, Fig.~S1 \cite{SI}), and we find that the power-law scaling with $n$ exponent $\mathcal{\gamma}$ and prefactor $a$ fits the RMSE better. The equation takes the form:
\begin{equation} \label{RMSE}
    \text{RMSE}(M_{n\times n},M_{30\times30}) = \sqrt{\frac{\sum_g (M_{n\times n,g} - M_{30\times30,g})^2}{\mathcal{N}}} \approx an^{\gamma}.
\end{equation} 
where $\mathcal{N}=900$ is the number of points $g$ we evaluated in each phase diagram (30 evaluations for $h \in [0,6]$ and 30 evaluations for $J_2 \in [0,6]$ with a step size of 0.2 in each variable).  The computed RMSEs and fitted power-law curves are shown in Fig.~\ref{tri_finite_size_exponential_scaling} while . Empirically, the RMSE diminishes following a power law scaling with the exponent $\mathcal{\gamma} = -1.27(4) $ for the triangular, and $\mathcal{\gamma} = -1.34(4) $ for the Shastry-Sutherland lattice. We note that the size of the $boundary$ scales as $O(n)$ while the size of the $bulk$ scales as $O(n^2)$. If the RMSE had arisen strictly from the boundary effects, it would diminish following the proportion of $boundary/bulk$  $ \sim O(1/n)$. However, $\mathcal{\gamma} < -1 $ signifies a faster drop off of RMSE as compared to $1/n$, which could be because of enhanced correlations between the various spins subject to the Hamiltonian.

A more rigorous analysis of finite-size scaling \cite{Goldenfeld,Khalid2022finitesize} around each critical point could yield a careful analysis of the required lattice size for target fidelity for every phase transition, our result based on an overall RMSE demonstrates that a 15$\times$15 spin grid is already obtaining results close to the much larger 30$\times$30 grid. Based on this trend, we expect that finite size effects in $M$ will diminish quickly with the size of the lattice, indicating that lattices of only a few hundred spins may diminish the errors sufficiently to achieve a realistic "bulk", and therefore meaningful results for comparison with experiments which probe bulk properties, such as diffraction and heat-capacity. This suggests that quantum processors with hundreds of qubits, achievable within the noisy intermediate-scale quantum era \cite{Preskill2018quantum}, may be capable of meaningful applications for materials science applications.
%
%%%%%%%%%%%%%%%%%%%%%%%%%%%%%%%%%%%%%%%%%%%%%%%%%%%%%%
\section{Quantum Approximate Optimization Algorithm} 
Quantum computers offer a route to overcoming issues associated with identifying ground states through conventional methods.  One approach to address these problems uses the quantum approximate optimization algorithm, which was originally designed to find approximate solutions to difficult combinatorial optimization problems \cite{farhi2014quantum} that are often expressed in terms of Ising Hamiltonians \cite{Lucas2014qubo}. Empirical performance of QAOA has been characterized for a variety of combinatorial problems \cite{Lotshaw2021empirical,zhou2020quantum,Bartschi2020MaxkCover,Pontus2020tail,Harwood2021routing} and this has also led to generalizations \cite{hadfield2019quantum,Patti2021nonlinearqaoa,Herrman2021ma,farhi2017hardware,Gupta2020WarmStart,zhu2020adaptqaoa} that have been applied to preparing chemical ground states \cite{Kremenetski2021qaoachemistry} as well as ground state preparation for one-dimensional \cite{Ho2019variationalstateprep,Matos2021stateprep} and two-dimensional \cite{Sun2022PRHQAOA} quantum spin models in theory and experiment \cite{pagano2020quantum}.  
\par
To formulate our Ising problems in a structure that is suitable for QAOA, we first express the Ising Hamiltonian (\ref{General Ising Model}) in terms of a quantum Hamiltonian operator
\begin{equation} \label{H} H = J_1 \sum_{(i,j) \in \mathrm{NN}} Z_iZ_j + J_2\sum_{(i,j) \in \mathrm{NNN}} Z_iZ_j + h\sum_{i=1}^N Z_i. \end{equation}
Here the $N$ spins $s_i \in \{+1,-1\}$ in (\ref{General Ising Model}) are encoded into the eigenvalues of the Pauli $Z$ operators, with $Z_i \vert z_i\rangle = s_i \vert z_i\rangle$, where $z_i \in \{0,1\}$ and $s_i = 1-2z_i$. The set of all spin values is then encoded into a computational basis state $\vert \bm z \rangle = \bigotimes_{i=1}^N \vert z_i\rangle$. Each $\vert \bm z \rangle$ is an energy eigenstate of $H$ with the energy eigenvalue of the corresponding classical spin problem,
\begin{equation} H\vert \bm z \rangle = \mathcal{H}(\bm z) \vert \bm z \rangle,\end{equation}
where $\mathcal{H}(\bm z)$ comes from (\ref{General Ising Model}) taking $s_i = 1-2z_i$ for each component $\vert z_i\rangle$ in the total basis state $\vert \bm z\rangle$. This gives an encoding of the Ising spin problem (\ref{General Ising Model}) that is useful for QAOA, where we will sample eigenstates $\vert \bm z \rangle$ to try to identify the ground state of the Ising problem. 
\par
\begin{figure*}
\includegraphics[width=12cm,height=20cm,keepaspectratio]{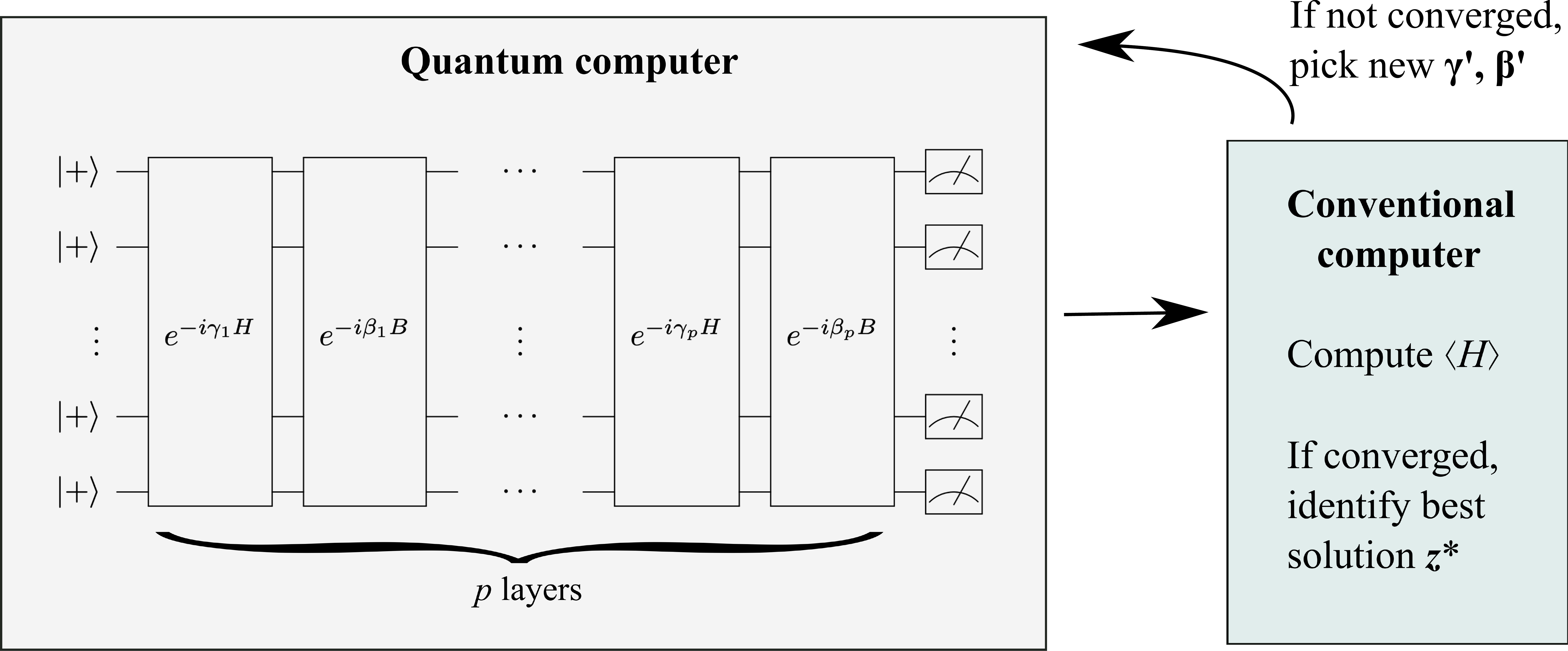}
\caption{Quantum-classical optimization loop for QAOA. For a given set of parameters $\bm \gamma, \bm \beta$, a quantum computer generates and measures states $\vert \psi_p(\bm \gamma, \bm \beta)\rangle$.  The measurements are sent to a conventional computer to compute $\langle H \rangle$ and check its convergence. If $\langle H \rangle$ is not converged, then an optimization routine selects new $\bm \gamma'$ and $\bm \beta'$ for the quantum computer. If $\langle H \rangle$ is converged, then the algorithm terminates and the final solution is the measured result $\bm z^*$ that minimizes the energy.}
\label{QAOA schematic}
\end{figure*}

To find solutions, QAOA uses a quantum state prepared with $p$ layers of unitary evolution, where each layer alternates between Hamiltonian evolution under the Ising Hamiltonian $H$ and under a ``mixing'' Hamiltonian $B = \sum_i X_i$
\begin{equation} \label{QAOA} \vert \psi_p(\bm \gamma, \bm \beta) \rangle = \left(\prod_{l=1}^p e^{-i \beta_l B}e^{-i \gamma_l H}\right) \vert \psi_0\rangle\end{equation}
where the initial state $\vert \psi_0\rangle = 2^{-N/2} \sum_{\bm z} \vert \bm z \rangle$ is the ground state of $-B$ represented in the computational basis.  The parameters $\bm \gamma = (\gamma_1,...,\gamma_p)$ and $\bm \beta = (\beta_1,...,\beta_p)$ are typically chosen to minimize the expectation value of the energy $\langle H \rangle$, though other objectives have also been studied  \cite{Barkoutsos2020CVaR,Kremenetski2021qaoachemistry,LiLi2020Gibbs}.  The minimization is typically accomplished using a quantum-classical hybrid feedback loop, shown schematically in Fig.~\ref{QAOA schematic}. For a given set of parameters $\bm \gamma$ and $\bm \beta$, a set of states $\vert \psi_p(\bm \gamma, \bm \beta) \rangle$ is prepared and measured by a quantum computer. The measurement results are sent to a conventional (classical) computer to compute the classical objective function.  If the objective function is not converged relative to previous evaluations, then the conventional computer uses an optimization routine to select new parameters $\bm \gamma', \bm \beta'$.  The process is repeated until convergence to a minimal objective with optimized parameters $\bm \gamma^*, \bm \beta^*$.  The final result is taken as the measurement result $\vert \bm z^*\rangle$ that gives the lowest energy $\mathcal{H}(\bm z^*)$.  In the best case, $\bm z^* = \bm z_\mathrm{ground}$ is a ground state, while more generally $\bm z^*$ may be a low-energy state that is an approximate solution to the problem.  
\par
An analytic proof has demonstrated that QAOA can prepare an exact ground state $\vert \bm z_\mathrm{ground}\rangle$ of Ising Hamiltonians as $p\to \infty$ \cite{farhi2014quantum,Lucas2014qubo}.   Apart from the formal proof of convergence at large $p$, there has been significant interest in applying QAOA at small $p$, where approximations exceeding conventional lower bounds have been observed in simulations \cite{Lotshaw2021empirical,crooks2018performance} and predicted for large problems in certain contexts \cite{Basso2022QAOAadvantage}.  Realizing such advantages on devices with hundreds of qubits or more is an important topic of ongoing research as quantum computing technologies continue to develop. 
\par
For the materials problems we consider, we are interested in preparing the ground states of the Ising Hamiltonian.  We compute exact ground states for our unit cells in Fig.~\ref{lattices} by evaluating all eigenvalues of the Hamiltonian using Eqs.~(\ref{General Ising Model}) and (\ref{H}) to identify the lowest energy state. For some phases the number of ground states is $N_\mathrm{ground}>1$ due to degeneracies, while for other phases there is a single ground state $N_\mathrm{ground}=1$, as pictured in Supplemental Information \cite{SI}. To assess QAOA performance, we compute the average ground state probability
\begin{equation} \label{Pground} \overline{P}_\mathrm{ground} = \frac{1}{N_\mathrm{ground}}\sum_{\bm z_\mathrm{ground}} P(\bm z_\mathrm{ground}),\end{equation}
where the sum contains a single term in the case of a non-degenerate ground state or multiple terms in the degenerate case. Analytically, the probabilities are given by the Born rule $P(\bm z) = \vert \langle \bm z \vert \psi_p(\bm \gamma, \bm \beta)\rangle|^2$, while for experiments on a quantum computer they are given by the frequencies of measurement results, $P(\bm z) = N(\bm z)/N_\mathrm{tot}$, where $N(\bm z)$ is the number of times $\vert \bm z \rangle$ was measured and $N_\mathrm{tot}$ is the total number of measurements. If QAOA identifies a ground state then $\vert z^* \rangle = \vert z_\mathrm{ground}\rangle$ and $\overline{P}_\mathrm{ground} > 0$, while if QAOA only finds sub-optimal solutions then $\overline{P}_\mathrm{ground}=0$. 
\par
Ground state preparation is a goal specific to the materials problem context we are interested in here.  This is, importantly, a departure from the standard goal of QAOA in the context of approximate combinatorial optimization, where the goal is to find approximate solutions that are not necessarily the ground states. While QAOA is not expected to efficiently find exact ground states for generic NP-hard optimization problems, it may still prove useful for finding ground states of specific structured problems such as materials problems on a lattice similar to those we explore here \cite{Ho2019variationalstateprep,Matos2021stateprep,pagano2020quantum,Sun2022PRHQAOA}.
\subsection{Numerical simulations of ideal QAOA}\label{simulations}
We use numerical calculations to assess the theoretical performance of QAOA for ground state preparation.  These demonstrate the ideal performance of QAOA in exact pure state calculations that use matrix multiplication to evaluate (\ref{QAOA}). This gives an ideal baseline for later comparison against results from a noisy quantum computer, where errors lead to mixed states with degraded performance.  We use $p=1$ QAOA layers throughout this section and our results.
\par
To identify QAOA states for our Ising problems we must determine optimized QAOA parameters $\gamma_1^*$ and $\beta_1^*$.  We choose regions to evaluate parameters in determining $\gamma_1^*$ and $\beta_1^*$ as follows.  QAOA is periodic when $\beta_1 \to \beta_1 \pm \pi$ \cite{zhou2020quantum}, hence we consider $-\pi/2 \leq \beta_1 \leq \pi/2$, which gives all unique $\beta_1$ up to symmetries.  The periodicity of the $\gamma_1$ parameter is more complicated, as it depends on the Hamiltonian in $\exp(-i \gamma_1 H)$ in (\ref{QAOA}). Here we focus on $\gamma_1$ intervals near the origin and dependent on the magnitude of the Hamiltonian terms, which has been highly successful in previous work \cite{shaydulin2022weightedtransfer}.  The basic idea is that the QAOA unitary $\exp(-i \gamma_1 H)$ changes at varying speeds with respect to $\gamma_1$, depending on the Hamiltonian coefficients $J_1$, $h$, and $J_2$. When the Hamiltonian coefficients increase, then $\gamma_1$ should decrease to obtain a similar unitary. The rate at which the unitary changes with respect to $\gamma_1$ is related to the average magnitude of the Hamiltonian coefficients
\begin{equation} \label{iota}\iota = \frac{Nh + J_1E_\mathrm{NN} + J_2E_\mathrm{NNN}}{N + E_\mathrm{NN} + E_\mathrm{NNN}},\end{equation}
where $E_\mathrm{NN}$ is the number of nearest-neighbor interactions and $E_\mathrm{NNN}$ is the number of next-nearest-neighbor interactions.  Previous work on generic Ising Hamiltonians with $h=0$ has shown that high quality solutions are obtained at small $\gamma_1$ with an empirical scaling of optimized parameters similar to $\gamma_1^* \sim 1/\iota$. The scaling $1/\iota$ compensates for the varying rates of evolution that are present for varying choices of the Hamiltonian, and also limits the interval of $\gamma_1$ values that are explored, simplifying the optimization \cite{shaydulin2022weightedtransfer}.   Based on these ideas we choose $\gamma_1$ in the interval $-0.55\times\pi/\iota \leq \gamma_1 \leq 0.55\times\pi/\iota$. 
\par
\begin{figure*}
   \begin{subfigure}{0.55\textwidth}
    \includegraphics[width=100cm,height=4.25cm,keepaspectratio]{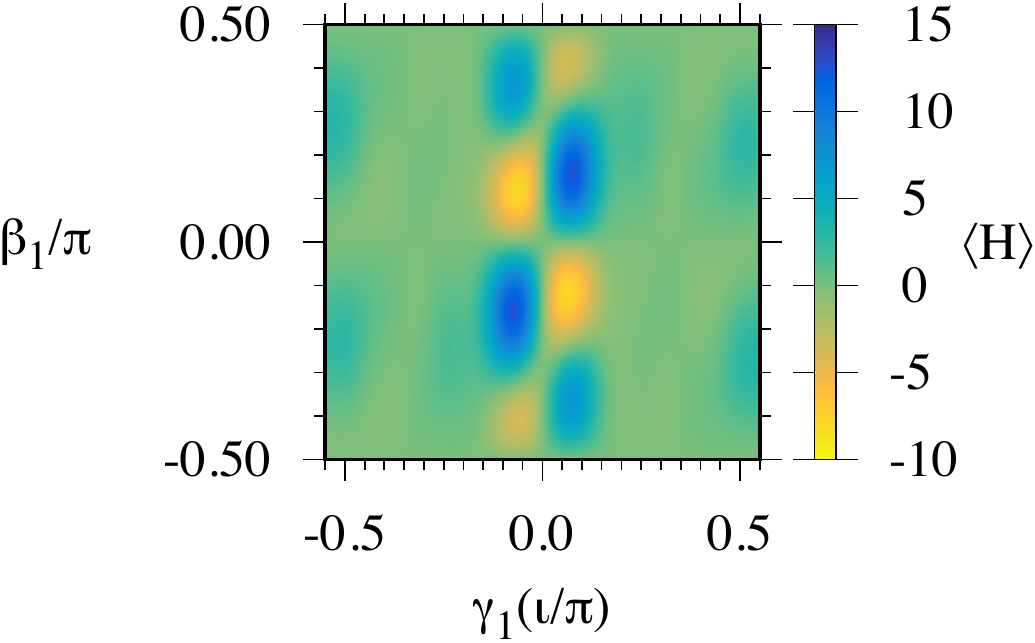}
    \caption{}
    \end{subfigure}
    \begin{subfigure}{0.45\textwidth}
    \includegraphics[width=100cm,height=4.25cm,keepaspectratio]{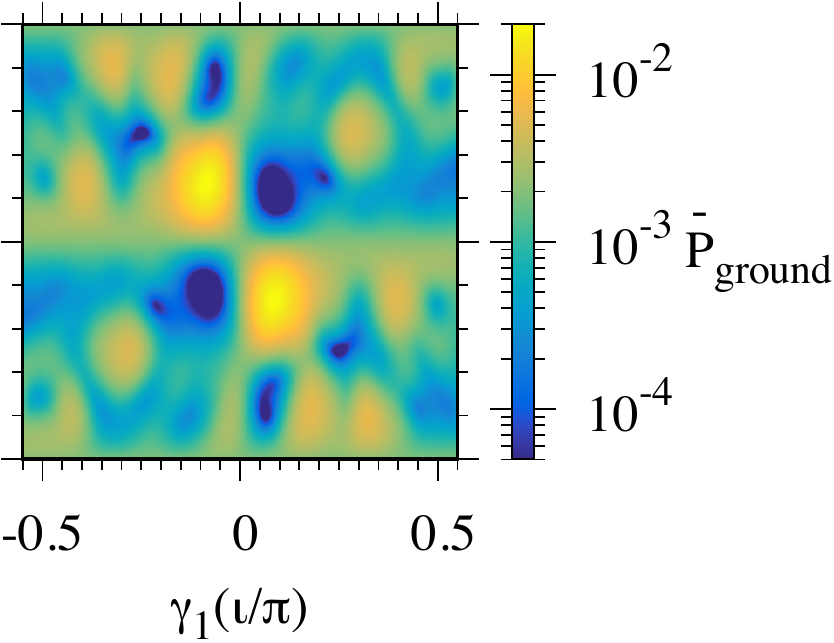}
    \caption{}
    \end{subfigure}
\caption{Numerical simulations of the (a) average energy $\langle H \rangle$ and (b) average ground state probability $\overline{P}_\mathrm{ground}$ with varying choices of QAOA parameters $\gamma_1$ and $\beta_1$, for the Shastry-Sutherland unit cell with $J_1 = 1, J_2 = 3.7,$ and $h = 1.4$ (Sec.~3\ref{simulations}). Each plot has the same range of $\beta_1$ and $\gamma_1$; the color scales are reversed in (a) and (b) so that small $\langle H \rangle$ and large $\overline{P}_\mathrm{ground}$ are each represented by bright colors.}
\label{contours}
\end{figure*}

We show an example of how the energy expectation value and average ground state probability depend on the choice of parameters in Fig.~\ref{contours} for an example with the Shastry-Sutherland unit cell with Hamiltonian coefficients $J_1 = 1, J_2 = 3.7,$ and $h = 1.4$ (similar patterns are observed in sample calculations using other choices of Hamiltonian coefficients and also for the triangular unit cell).  There are two regions in Fig.~\ref{contours}(a) with optimized $\langle H \rangle$ in yellow.  The ground state probabilities in Fig.~\ref{contours}(b) are also relatively large near the $\gamma_1^*$ and $\beta_1^*$ that optimize $\langle H \rangle$. 
\par
We have found in searches over much larger $\gamma_1$ intervals that the local optima for $\overline{P}_\mathrm{ground}$ and $\langle H \rangle$ do not always approximately align as in Fig.~\ref{contours}, which can lead to poor $\overline{P}_\mathrm{ground}$ at optimized $\langle H \rangle$ in these larger intervals.  However, this does not appear to be a prevalent issue for the smaller $\iota$-dependent $\gamma_1$ intervals in cases we have looked at. The results are somewhat sensitive to the specific choice of $\gamma_1$ interval, however, our choice of $-0.55\times\pi/\iota \leq \gamma_1 \leq 0.55\times\pi/\iota$ gives satisfactory results across the varying lattices.
\par
To identify optimized parameters, we perform a grid search over $\gamma_1$ and $\beta_1$ for each Hamiltonian considered.  We evaluate the QAOA states in (\ref{QAOA}) on 201 evenly spaced intervals with $-\pi/2 \leq \beta_1 \leq \pi/2$ and over 300 evenly spaced intervals with $-0.55\times\pi/\iota \leq \gamma_1 \leq 0.55\times\pi/\iota$ for a total of $201\times 300 = 60,300$ grid evaluations for each Hamiltonian.  This approach gives optimal parameters in our intervals up to coarse graining in the grid search.  We select parameters $\gamma_1^*$ and $\beta_1^*$ that optimize $\langle H \rangle$.  
\par
The optimized parameters $\gamma_1^*$ and $\beta_1^*$ do not necessarily give the optimal ground state probabilities that are possible from QAOA.  The reason is that the average energy optimization accounts for energies and probabilities of all states, which together may yield low energies at parameter choices that are not optimal for the ground states alone \cite{Lotshaw2021empirical}. To assess performance, we further compare our parameter choices against parameters $\gamma_1^{*\prime}$ and $\beta_1^{*\prime}$ that directly optimize $\overline{P}_\mathrm{ground}$. The direct optimization of $\overline{P}_\mathrm{ground}$ is used here for benchmarking purposes and is not a realistic approach for large problems where the ground states are unknown. For our small problems, the comparison gives an idea of how the ground state probabilities from a standard optimization of $\langle H \rangle$ compare against the best ground state probabilities that could be obtained by QAOA in our setup.
\subsection{Quantum computations of QAOA}\label{qcomp}
We next investigate the performance of QAOA using the Quantinuum H1-2 quantum computer.  H1-2 contains trapped-ion qubits and uses lasers to implement gates on these qubits.  Typical error rates are reported as $3.5 \times 10^{-3}$ for two-qubit gates and $1 \times 10^{-4}$ for single-qubit rotation gates \cite{Honeywell_device_spec}. In addition to the device H1-2, we also us the H1-1E device "emulator" to simulate noisy device behavior. This gives results that approximately correspond to expected device behavior while avoiding the financial expense and wait times that are associated with running the device.  The emulator models a variety of device-specific noise processes for the H1-class computers, including depolarizing noise, leakage errors, crosstalk, dephasing in transport, and qubit idling errors \cite{Honeywell_emulator_spec}.
\par 
We test QAOA on the H1-2 using the QCOR software stack \cite{Mintz2020QCOR}.  The QCOR stack translates the series of unitary operators expressing QAOA into quantum circuits for H1-2; see Supplemental Information \cite{SI}, Appendix B, for details. The QCOR program used for submitting jobs to the device as well as our calculations are available online, cf. "Data Accessibility".
\par
Furthermore, modern quantum computers are known to be affected by state preparation and measurement (SPAM) errors as well as gate infidelities from a variety of physical sources.  We assessed SPAM errors expected in our quantum computations using the device emulator, with details in Supplemental Information Appendix C \cite{SI}. The probability to observe no error in circuits we tested was approximately 96\%, with errors distributed approximately uniformly across qubits. We account for these errors using an independent bit-flip model and associated SPAM matrix $\tilde P$, which transforms an ideal set of measurement results to the expected noisy set of results.  The inverse matrix $\tilde P^{-1}$ can then be applied to our noisy measurements from the quantum computer to approximately correct for SPAM errors. A technical issue arises in that the mitigated measurement probabilities can sometimes be negative, due to the approximate nature of the mitigation scheme.  This leads to a second mitigation scheme that additionally sets all negative probabilities to zero and renormalizes so the total probability is one. We use each of these approaches to attempt to correct the small SPAM errors we expect from the quantum computer, as described in detail in Supplemental Information \cite{SI}. 
%
%%%%%%%%%%%%%%%%%%%%%%%%%%%%%%%%%%%%%%%%%%%%%%%%%%%%%%
\section{Results}
In this section we consider the results from QAOA applied to the materials lattices of Fig.~\ref{lattices}.  We take $J_1=1$ as the unit of energy and analyze the success of QAOA in preparing ground states at variable $h$ and $J_2$, first in numerical simulations (Sec.~3\ref{simulations}) and then in quantum computations on a trapped-ion quantum computer (Sec.~3\ref{qcomp}).
\subsection{Ground state measurement probabilities} \label{simulation results}
\begin{figure*} 
   \begin{subfigure}{0.55\textwidth}
    \includegraphics[width=6.7cm,height=10cm,keepaspectratio]{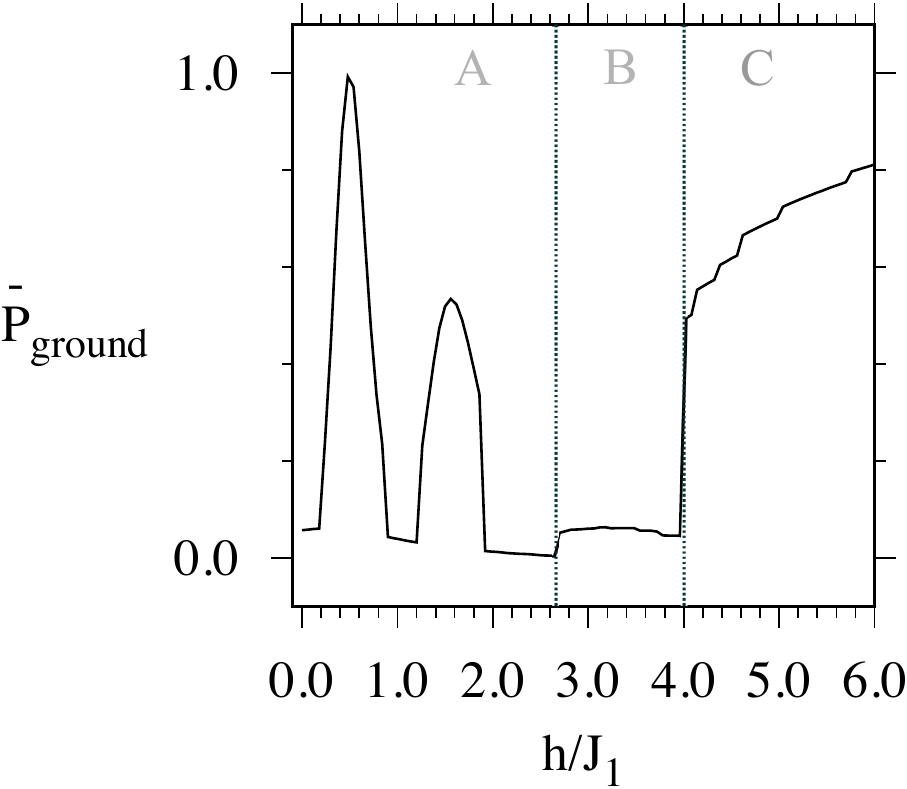}
    \caption{}
    \end{subfigure}
    \begin{subfigure}{0.45\textwidth}
    \includegraphics[width=4.75cm,height=10cm,keepaspectratio]{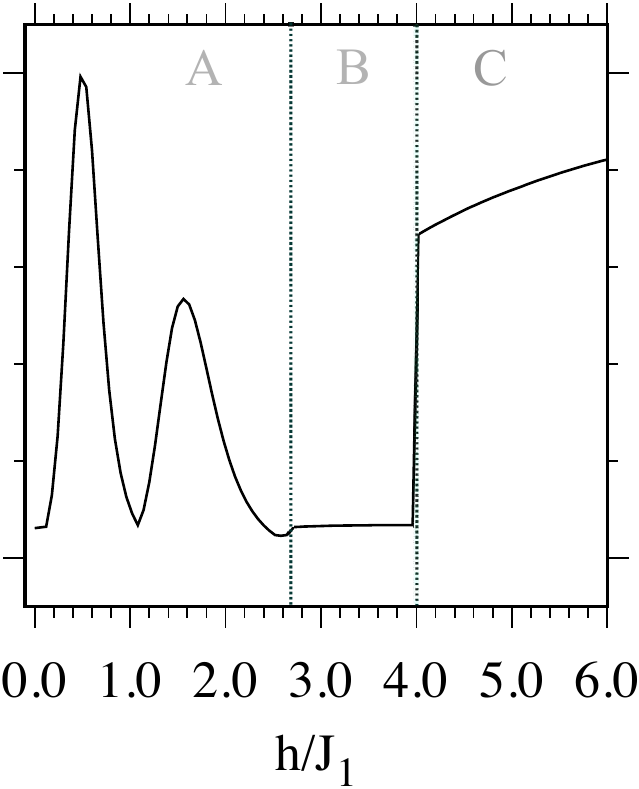}
    \caption{}
    \end{subfigure}
\caption{The square unit cell ground state probabilities when (a) optimizing $\langle H \rangle$ and (b) optimizing $\overline{P}_\mathrm{ground}$ as described in Sec.~3\ref{simulations}. The $\overline{P}_\mathrm{ground}$ ranges are identical in each figure. Phases A, B, C refer the anti-ferromagnetic, $M$=7/9, and ferromagnetic phases of Fig.~\ref{lattices} respectively, with vertical dotted lines showing the phase boundaries.}
\label{square phase}
\end{figure*}

We first consider theoretical probabilities to measure the ground state with QAOA and how these vary for different parameter choices in the Hamiltonian.  We begin with the simple square lattice in Fig.~\ref{lattices}(a,d), which does not exhibit frustration as there are no triangles in the interaction graph.  The probability to measure the ground state for varying $h$ is shown in Fig.~\ref{square phase}.  Fig.~\ref{square phase}(a) shows the probabilities obtained from optimizing the standard objective $\langle H\rangle$ while Fig.~\ref{square phase}(b) shows the best-case results based on optimizing $\overline{P}_\mathrm{ground}$, as described in  Sec.~3\ref{simulations}. The probabilities in each case are similar, demonstrating that optimizing $\langle H \rangle$ is nearly as successful in increasing the ground state probability as a direct optimization.
\par
The average ground state probability shows distinct behaviors for each of the three ground states at varying $h$, visually separated by dotted lines.  In the anti-ferromagnetic ground state at small $h$, the probability $\overline{P}_\mathrm{ground}$ approximately oscillates between $h=0$ and $h=2$, with small probabilities observed near integer values of $h$ and larger probabilities near $h=1/2$ and $h=3/2$. The $M=7/9$ ground state with $8/3 < h < 4$ has a near-constant probability of $\approx 0.06$.  At $h=4$ the ground state becomes ferromagnetic and $\overline{P}_\mathrm{ground}$ increases significantly, with monotonic increases at larger $h$.
\par
We rationalize the varying success probabilities in the figure as attributable to structures of the ground states at varying $h$ and the interplay with the structure of the QAOA state in (\ref{QAOA}).  We show in Supplemental Information \cite{SI} Appendix D that QAOA can exactly prepare the ferromagnetic ground state when $h \gg J_1$ for arbitrary lattice sizes, based on the fact that $\exp(-i \gamma H) \approx \exp(-i \gamma h \sum_{i=1}^N Z_i)$ in this limit.  This is consistent with the behavior in the figure, where $\overline{P}_\mathrm{ground}$ increases monotonically with $h$ for the ferromagnetic ground state at $h \geq 4$. We further show in Supplemental Information \cite{SI} Appendix E how the anti-ferromagnetic ground state probability is maximized at $h=0.5$, and we devise large $\gamma_1$ parameters that can further improve these results (We did not include larger $\gamma_1$ parameters in our numerical searches as this can lead to poor $\overline{P}_\mathrm{ground}$ at parameters that optimize $\langle H \rangle$ for the frustrated lattices, as remarked in Sec.~3\ref{simulations}).  However, the mechanism for anti-ferromagnetic ground state preparation here depends on the specific choice of the $3\times3$ lattice, and it is not clear how QAOA will behave for other lattice sizes.  For the M=7/9 phase the QAOA state is more complicated, as it is a superposition of many basis states that depends on the optimized parameters, and we do not have an analytic account for this behavior.  The optimized parameters that create each QAOA state in the square-lattice phase diagram are shown in Supplemental Information Appendix F \cite{SI}.
\par
\begin{figure*}
    \begin{subfigure}{0.5\textwidth}
    \includegraphics[width=6.cm,height=6.5cm,keepaspectratio]{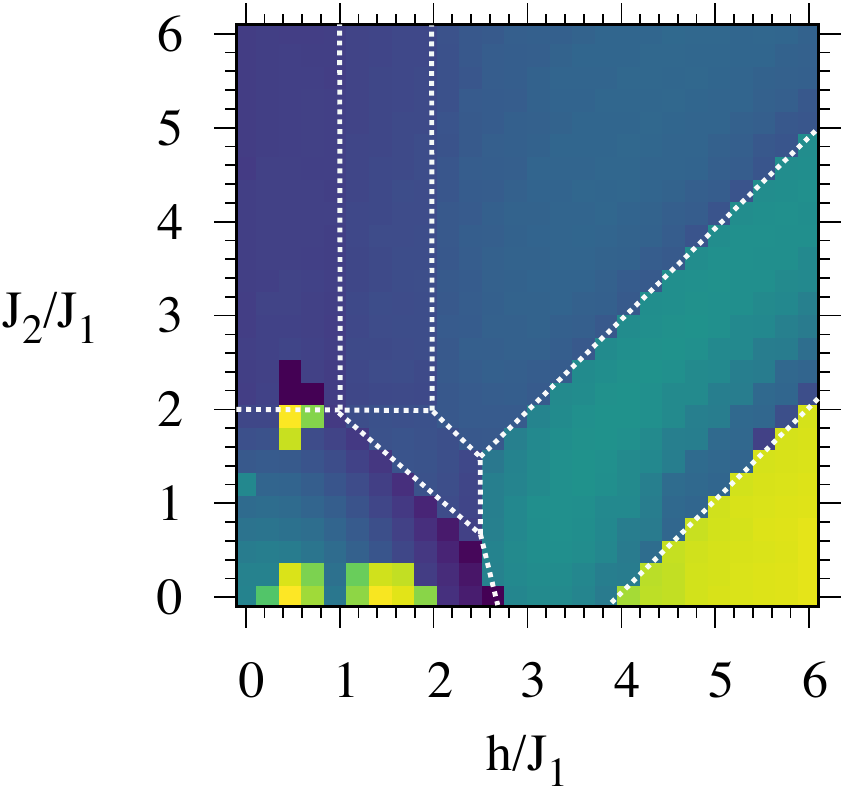}
    \caption{}
    \end{subfigure}
        \begin{subfigure}{0.5\textwidth}
    \includegraphics[width=7.cm,height=8cm,keepaspectratio]{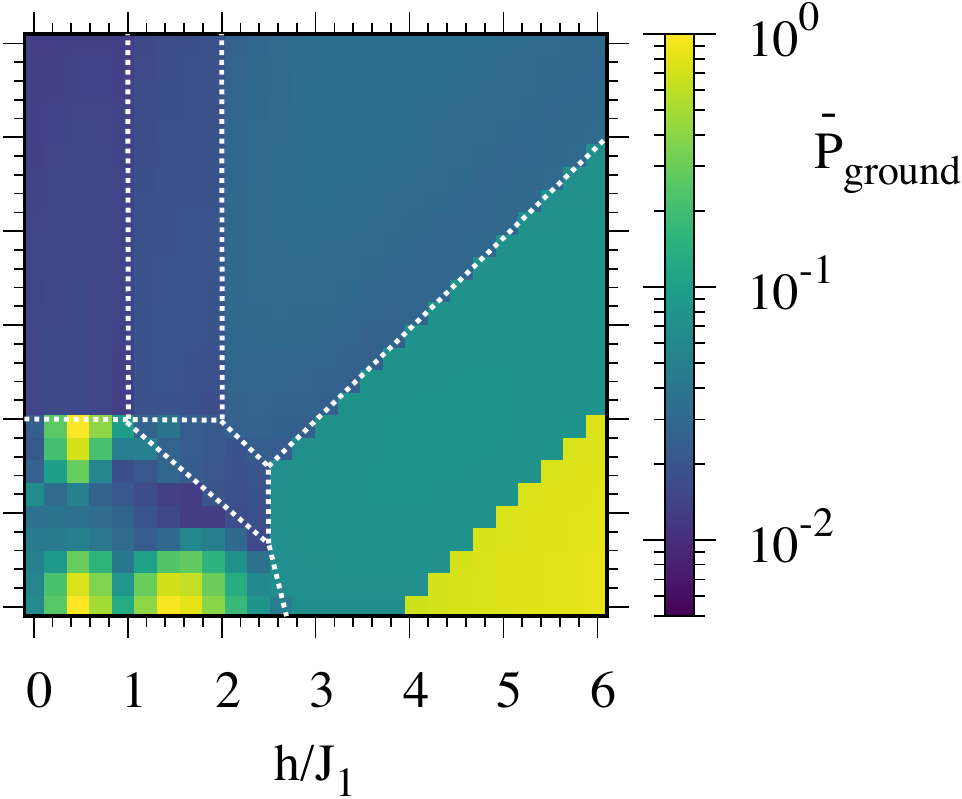}
    \caption{}
    \end{subfigure}
\caption{Shastry-Sutherland unit cell ground state probabilities when (a) optimizing $\langle H \rangle$ and (b) optimizing $\overline{P}_\mathrm{ground}$ as described in Sec.~3\ref{simulations}. The ranges for $J_2/J_1$ and $\overline{P}_\mathrm{ground}$ are identical in each figure.}
\label{SS phase}
\end{figure*}
\begin{figure*}
   \begin{subfigure}{0.5\textwidth}
    \includegraphics[width=6.cm,height=10cm,keepaspectratio]{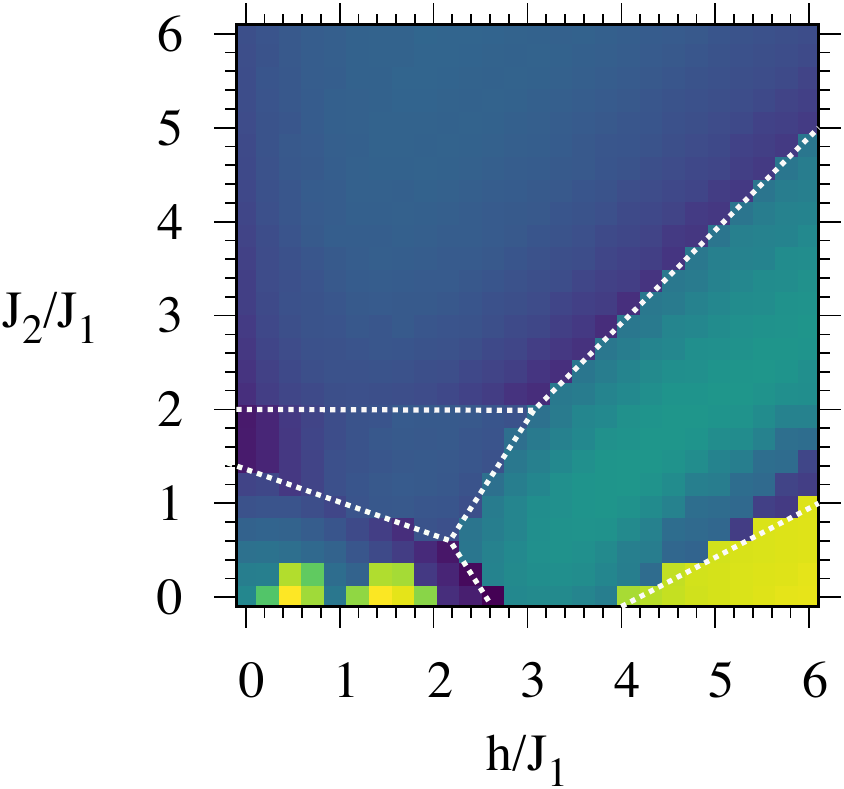}
    \caption{}
    \end{subfigure}
        \begin{subfigure}{0.5\textwidth}
    \includegraphics[width=7cm,height=10cm,keepaspectratio]{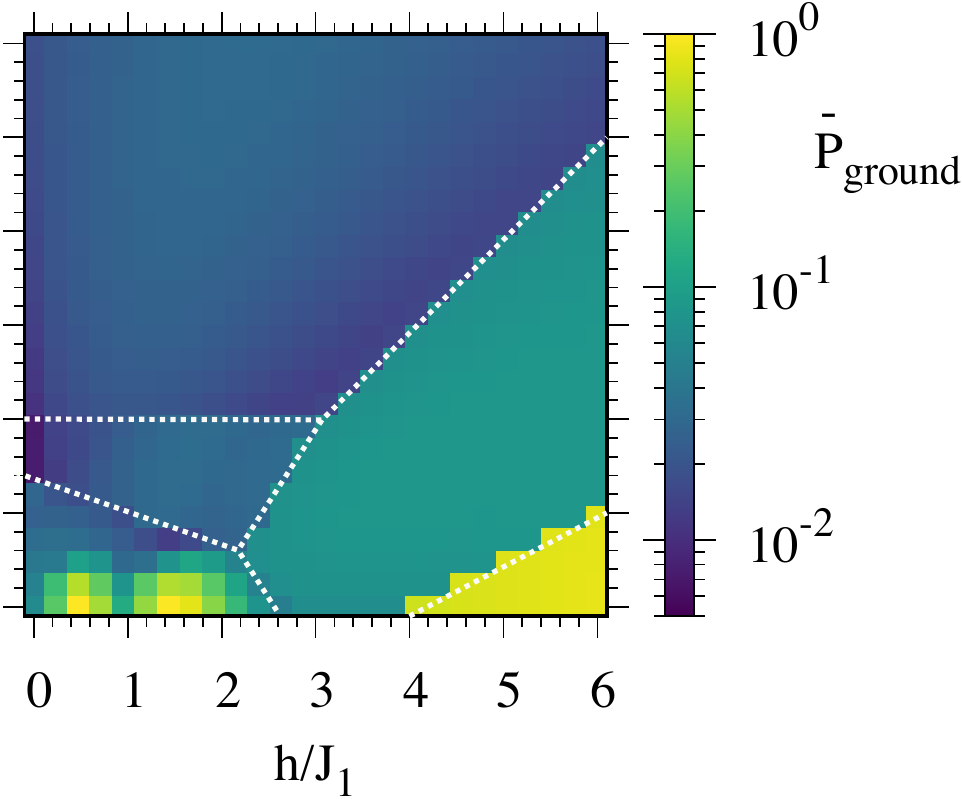}
    \caption{}
    \end{subfigure}
    \caption{Triangular unit cell ground state probabilities when (a) optimizing $\langle H \rangle$ and (b) optimizing $\overline{P}_\mathrm{ground}$ as described in Sec.~3\ref{simulations}. The ranges for $J_2/J_1$ and $\overline{P}_\mathrm{ground}$ are identical in each figure.}
\label{triangle phase}
\end{figure*}

Ground state probabilities $\overline{P}_\mathrm{ground}$ for the Shastry-Sutherland and triangular lattices are pictured in Figs.~\ref{SS phase} and ~\ref{triangle phase}, respectively. Ground state probabilities from optimizing $\langle H \rangle$ are presented in panels (a) while panels (b) show the best case probabilities from a direct optimization of $\overline{P}_\mathrm{ground}$ as described in Sec.~3\ref{simulations}. These probabilities show patterned behavior, with distinct probabilities $\overline{P}_\mathrm{ground}$ observed throughout most of each individual region A,B,..., with significant differences in $\overline{P}_\mathrm{ground}$ between different regions.  At small $J_2$, there are oscillations in the probability for preparing the anti-ferromagnetic ground states at small $h$, and large success probabilities for the ferromagnetic ground state at large $h$, as foreshadowed by results from the square lattice.  On the other hand, as the $J_2$ coupling increases, the triangular and Shastry-Sutherland lattices experience increased frustration, with competing interactions within the triangular motifs in Fig.~\ref{lattices}.  The average ground state probability decreases significantly as $J_2$ increases and frustration becomes dominant.  This is especially evident when $J_2 \gtrsim h$, for example in the top left of each of Figs.~\ref{SS phase}(b) and ~\ref{triangle phase}(b). The $\overline{P}_\mathrm{ground}$ are mostly uniform across $h$ and $J_2$ within each region, qualitatively similar to the nearly-uniform probability for the $M=7/9$ state at varying $h$ for the square lattice in Fig.~\ref{square phase}. Ground state probabilities are typically $\gtrsim$ 0.01, indicating that only $\lesssim 100$ measurements are expected to identify a ground state.   We now turn to computations on a trapped-ion quantum computer, to benchmark and assess performance of QAOA on a real quantum computing device.
\subsection{QAOA quantum computations}
Here we assess QAOA performance in preparing ground states on a trapped-ion quantum computer. Ultimately, our aim is to validate the idea that a current quantum computing technology is capable of preparing each ground state of a frustrated Ising Hamiltonian using QAOA.  An important first step is to assess whether optimized parameters from our theoretical calculations are also optimized for the device, or whether further optimization is needed to determine device-specific optimized parameters.
\subsubsection{Quantum computational performance with varying parameters}
\begin{figure*}
    \begin{subfigure}{0.5\textwidth}
    \includegraphics[width=8.25 cm,height=12 cm,keepaspectratio,trim={2cm 0 0 0},clip]{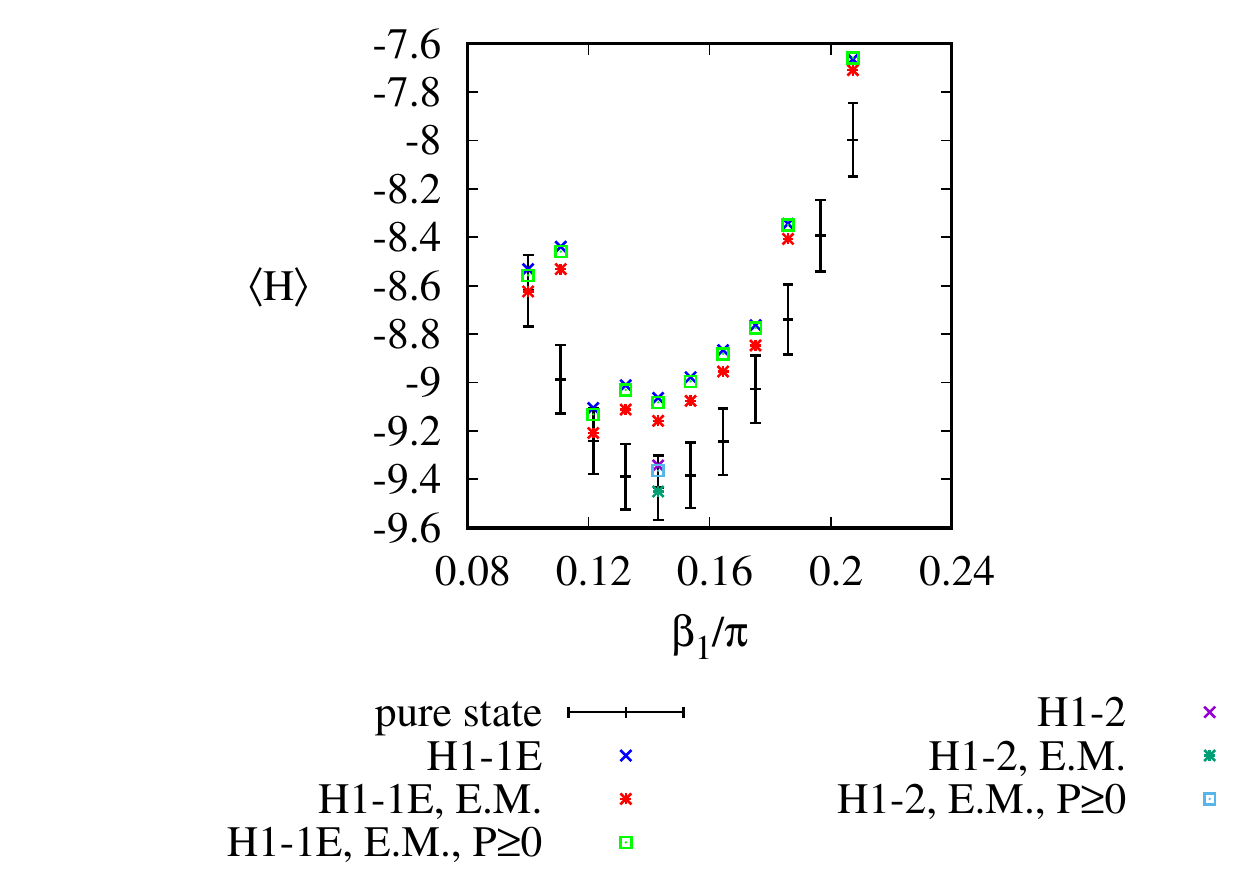}
    \caption{}
    \end{subfigure}
        \begin{subfigure}{0.5\textwidth}
    \includegraphics[width=8.25 cm,height=12 cm,keepaspectratio,trim={2cm 0cm 0cm 0},clip]{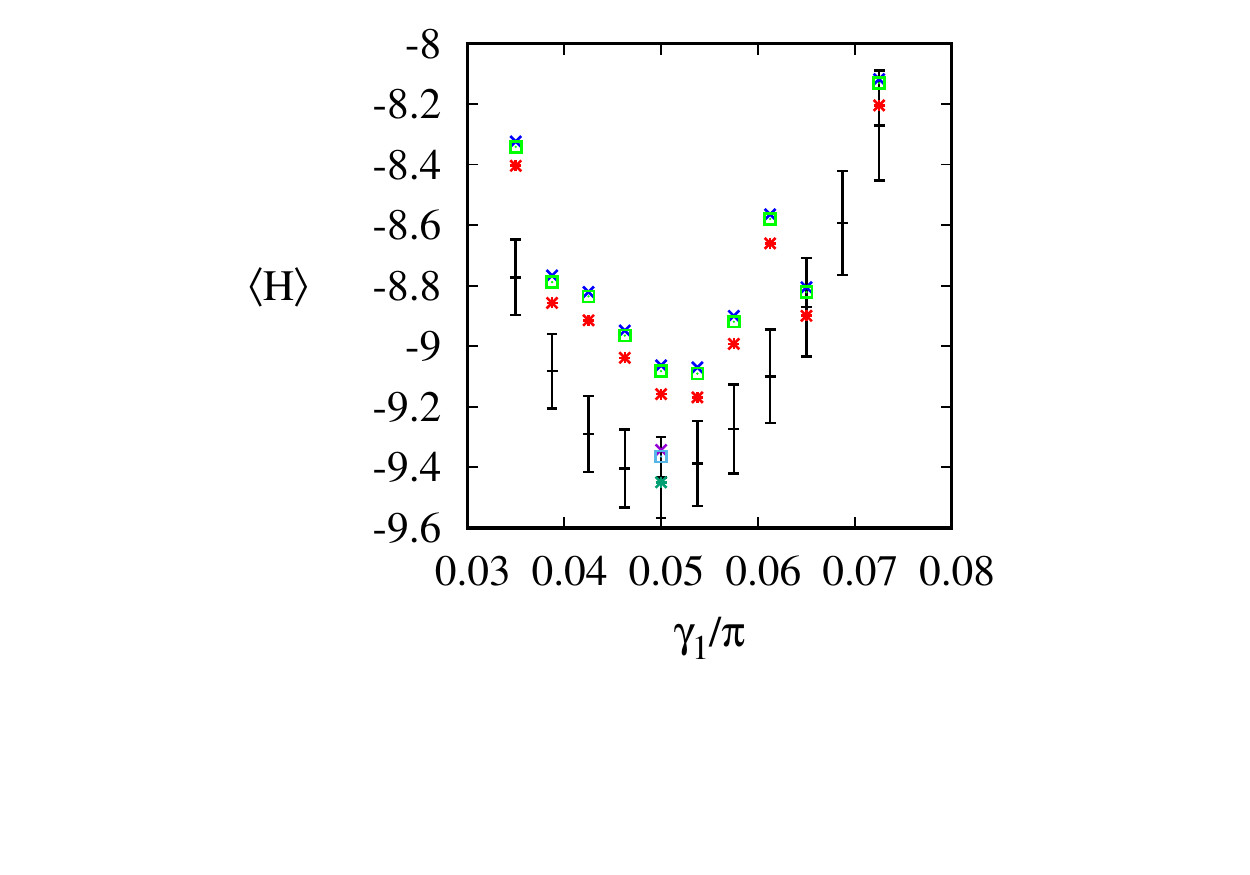}
    \caption{}
    \end{subfigure}
\caption{Angle sensitivity analysis for $h=2.48$ and $J_2 = 2.0$, with separate variations in (a) $\beta_1$ and (b) $\gamma_1$ about the ideal values from Table \ref{tab:quantum comp parameters}.  Black crosses show results from pure state calculations, with error bars denoting the standard error of the mean at 1000 shots (see text).  Data points showing results from the Honeywell emulator are denoted with "E" in H1-1E and results from the trapped-ion quantum computer are labeled H1-2. Data points labeled 'H1-1E' and 'H1-2' are raw data, labels "E.M." (error-mitigation) are with basic mitigation, and "E.M. P$\geq$0" are readout error mitigation that forces each probability $P(\bm z)$ to be $\geq 0$, as described in Sec.~3\ref{qcomp}.}
\label{sensitivity}
\end{figure*}

QAOA depends on the choice of parameters, as discussed in connection with Fig.~\ref{contours}.  To test whether our theoretical parameters also yield good performance in the device, we consider QAOA circuits evaluating a point in region E of the Shastry-Sutherland phase diagram Fig.~\ref{lattices}(e), with Hamiltonian coefficients and QAOA parameters shown in Table \ref{tab:quantum comp parameters}. The parameters correspond to a local minimum in $\langle H \rangle$, similar to the minima observed in Fig.~\ref{contours}.  We use the H1-1E device emulator to evaluate circuits at the optimized $\gamma_1$ and $\beta_1$ and circuits where either $\gamma_1$ or $\beta_1$ has been displaced from its optimal value, as shown in Fig.~\ref{sensitivity}. Black crosses in the figure indicate how $\langle H \rangle$ increases in pure state simulations as either of these parameters are varied individually.  Error bars denote the analytic standard error of the mean (S.E.M.) for $N_\mathrm{shots}=1000$ measurement shots, with S.E.M.$=\sqrt{(\langle H^2\rangle-\langle H\rangle^2)/N_\mathrm{shots}}$ calculated numerically from the pure states. If the quantum computations did not have any noise, then from the central limit theorem we would expect about two-thirds of the $\langle H \rangle$ from the quantum computer to be within these error bars.
\begin{table}[]
    \centering
    \begin{tabular}{c|c|c|c|c|c|c|c}
        region & degeneracy & $M$ & $J_2/J_1$ & $h/J_1$ & $\beta_1/\pi$ & $\gamma_1/\pi$ & $N_\mathrm{shots}$ \\
        \hline
        A & 1 & 1/9 & 0.240 & 1.440 & 0.750 & -0.507 & 400\\
        B & 4 & 1/9 & 3.840 & 0.480 & 0.112 & -0.048 & 1000\\
        C & 4 & 3/9 & 3.840 & 1.680 & 0.121 & -0.043 & 1000\\
        D & 2 & 3/9 & 1.680 & 1.920 & 0.131 & -0.056 & 400\\
        E & 4 & 5/9 & 2.000 & 2.480 & 0.143 & -0.050 & 1000\\
        F & 1 & 7/9 & 1.680 & 3.600 & 0.182 & -0.046 & 400\\
        G & 1 & 1.0 & 0.240 & 5.520 & 0.244 & -0.041 & 400\\
    \end{tabular}
    \caption{The parameters used for quantum computations with the Shastry-Sutherland lattice.  Here $J_2/J_1$ and $h/J_1$ are the Hamiltonian coefficients used in the calculations, $\gamma$ and $\beta$ are the QAOA parameters, and $N_\mathrm{shots}$ is the number of measurement shots taken on the quantum computer. }
    \label{tab:quantum comp parameters}
\end{table}
\par
The theoretical $\langle H \rangle$ can be compared against the device emulator, with data point labels in the figure beginning with ``H1-1E".  There are three sets of data points for the emulator; the first is direct output labeled ``H1-1E", the second includes SPAM error mitigation (Sec.~3\ref{qcomp}) in ``H1-1E, E.M.", the third includes a variation of the SPAM mitigation that additionally forces the mitigated probabilities to be $P \geq 0$ in ``H1-1E, E.M., P$\geq$0".  These emulated $\langle H \rangle$ are larger than the theoretical values and we attribute this to noise in the device emulator, which introduces errors that cause the energy to deviate from its ideal minimum value.  Despite these errors, the shape of the landscape is similar to our theoretical calculations, with best performance observed near the theoretical parameters that minimize $\langle H \rangle$, and energies that tend to increase away from these parameters. 
\par
We further validate that the H1-2 trapped-ion device itself is consistent with the emulator in the data points that begin with ``H1-2". These actually yield better energies than the emulator, and are within one standard error of the mean from our theoretical results. The results from the device and emulator indicate that the energy landscape as a function of the QAOA parameters $\gamma_1$ and $\beta_1$ is consistent between our theoretical calculations, the quantum device, and emulator.  We therefore proceed with our theoretically optimized parameters to evaluate success in ground state preparation using the quantum computer.
\subsubsection{Quantum calculations of ground states}
We now perform calculations on the Honeywell H1-2 quantum computer to analyze success probability in ground state preparation.  We consider points in each region of the Shastry-Sutherland lattice, using parameters listed in Table \ref{tab:quantum comp parameters} that correspond to local minima in $\langle H \rangle$, similar to the minima used to evaluate theoretical performance in Sec.~4\ref{simulation results}). We post-process the measurement results using the SPAM mitigation model with probabilities $P(\bm z) \geq 0$ (see Sec.~3\ref{qcomp}), to give a minor correction to the observed results that is designed to counteract this known source of error. 
\par
\begin{figure*}
    \centering
    \includegraphics[width=10cm,height=9cm,keepaspectratio,trim={0cm 0 0 0},clip]{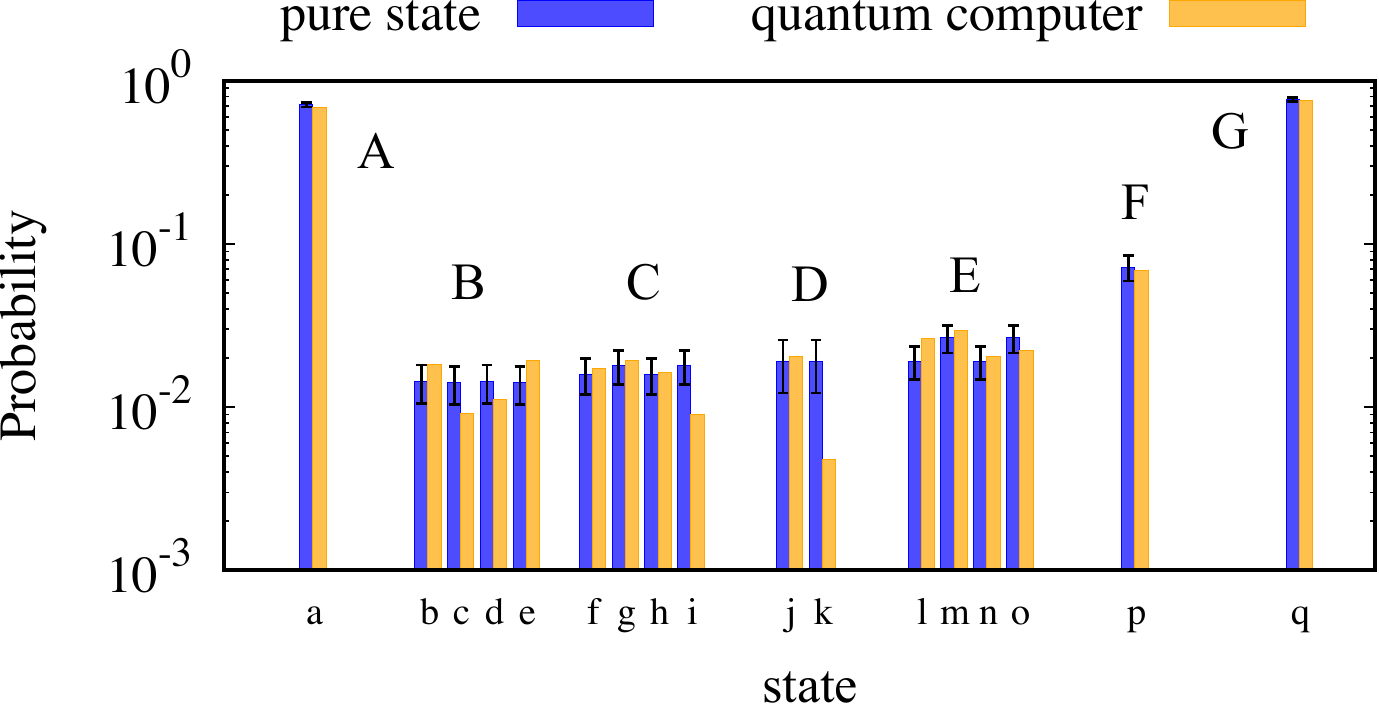}
\caption{Probabilities to observe each ground state from pure state simulations compared with observed frequencies estimated by quantum computations with the H1-2 device for each different phase (A-G) of the Shastry-Sutherland unit cell. Alphabetical labels ``a'', ``b'', etc., identify the different ground states in Fig.~\ref{fig:SS ground states}.}
\label{honeywell results}
\end{figure*}

Figure \ref{honeywell results} shows the ground state probabilities from quantum computations in comparison with ideal expectations from pure states.  The ground states are separated by regions A,B,... with markers a,b,... corresponding to the individual ground states pictured in Fig.~\ref{fig:SS ground states}.  The quantum computations succeed in observing each individual ground state in each region of the Shastry-Sutherland lattice, as seen by the positive probabilities in each state "a", "b",$\ldots$  
\par
For a closer comparison of probabilities, we plot error bars denoting the theoretical standard error of the mean S.E.M.=$\sqrt{P(\bm z)(1-P(\bm z))/N_\mathrm{shots}}$. The S.E.M. defines a range in which we expect about two-thirds of estimated $P_\mathrm{est}(\bm z) = N(\bm z)/N_\mathrm{shots}$ are expected to be found, where $N(\bm z)$ is the number of measurement results of a given ground state and $N_\mathrm{shots}$ is the total number of measurements in Table \ref{tab:quantum comp parameters}.  The probabilities from the quantum computation are largely consistent with the pure state results, with the majority of results within one S.E.M. from the ideal $P(\bm z)$, as expected in finite sampling to estimate the ground state probability. There are only two large deviations for states "k" and "i", which may be related to noise in the device.  The quantum computations succeed in preparing ground states with probabilities comparable to pure-state expectations.
%
%%%%%%%%%%%%%%%%%%%%%%%%%%%%%%%%%%%
\section{Conclusion}
We analyzed QAOA as an approach for preparing materials ground states on three types of Ising Hamiltonians with longitudinal magnetic fields, focusing on nine-spin unit cells as a starting size that is amenable to simulations and calculations on a trapped-ion quantum computer.  
\par
We applied QAOA to the nine-spin Ising unit cell problems to assess its success in ground state preparation. We found that the theoretical success probability depends significantly on the structure of the ground state, while it is mostly insensitive to the precise Hamiltonian parameters, which can vary within regions consistent with a fixed ground state. Each Hamiltonian yields a ferromagnetic ground state in the presence of large magnetic fields, and QAOA achieved large success probabilities for these relatively simple states. The probabilities for other types of ground states were more variable, and tended to decrease as next-nearest-neighbor couplings became stronger with associated frustration in the lattice. For all of these nine-spin states, we typically find success probabilities indicating that $\lesssim$ 100 measurements are expected to be necessary from an ideal quantum computer for these problem instances.
\par
To assess QAOA performance under realistic conditions we implemented the algorithm on a trapped-ion quantum computer. These quantum computations succeeded in observing each of the 17 ground states of the Shastry-Sutherland unit cell.  The quantum computations yielded ground state probabilities that were consistent with theoretical expectations based on pure states, indicating that noise was not a significant issue at the sizes and depths tested.  This suggests that calculations with current technology can likely be extended to greater QAOA depth parameters $p$, and to larger sizes as greater numbers of qubits become available. At greater depths and sizes we expect higher performance and more realistic results in comparison with the large size limit, respectively.
\par
While the ground states and associated phase diagrams for our nine-spin unit cells were found to have significant finite size effects relative to the large-size limit, the errors from finite size effects on the classically-calculated magnetization phase diagrams on $n \times n$ lattices up to $n=30$ was found to be suppressed rapidly with $n$, with small errors at $n=15$ indicating that only hundreds of spins may be necessary to reproduce large scale behavior.  This provides a baseline of hundreds of qubits for quantum computational experiments that seek to explain materials science problems, which may be accessible to near-term quantum computers in coming years.
\par
Assessing scaling of the ground state probability with size $N$ will be an essential aspect of extending this approach to larger sizes $N$. This includes numerical simulations to quantify how the ground state probabilities depend on the number of spins $N$ and number of QAOA layers $p$; previous works have shown $p \propto N$ maintains a large ground state probability $(\gtrsim 0.7)$ for simple models in different contexts \cite{biamonte2022depthscaling,Ho2019variationalstateprep}, but future work is needed to test scaling in the current model.  Benchmarking on quantum computers is also essential to understand how real noise processes effect scalability.
\par
Thus from our results we envision QAOA can be successfully applied to somewhat larger lattice problems as quantum computing technologies develop and larger number of qubits become available. These could be used for optimization of Ising lattice problems as we have here, with increasing sizes that potentially describe real bulk properties of materials in the $N\to\infty$ limit.  Additionally, the QAOA algorithm is general, with the application of a magnetic field, and hence could be explored for lattice problems which are NP-Hard \cite{barahona_1982}. But a more promising future direction leveraging the full benefit of this approach is to extend and modify QAOA to prepare ground states of quantum Hamiltonians such as the $XY$ and Heisenberg models, which can lead to a variety of quantum phenomena not captured in the Ising model, such as quantum spin glasses \cite{Heiko1997spinglass}, spin nematicity \cite{Reiss2021nematic}, Berzinski-Kosterlitz-Thouless states \cite{Hu2020BKTtransition, Kosterlitz1973transition} and long-range entangled states such as Dirac string excitations \cite{Jaubert2009spinice}, the likes of which exist in 2D frustrated quantum spin liquids and spin ice. Many of these topics are fiercely researched and are of considerable interest and importance for future quantum technologies and devices. Conventional numerical methods for understanding these states are hindered by the exponential size of the Hilbert space, making it difficult to generate a theoretical understanding of experimental observations.  QAOA or related generalizations \cite{hadfield2019quantum,Kremenetski2021qaoachemistry,Ho2019variationalstateprep,Matos2021stateprep,pagano2020quantum,Sun2022PRHQAOA} offer a potential route to overcome conventional computing bottlenecks.  Some successes along these lines have been observed in certain contexts, however, advances in methodology and quantum computing technologies are needed to extend these methods to complicated and larger-scale problems where quantum computational approaches may have a significant impact in understanding and developing materials for technological applications.
\enlargethispage{20pt}
\dataccess{Data and code from this study are available online at https://code.ornl.gov/5ci/dataset-simulations-of-frustrated-ising-hamiltonians-using-qaoa}
\aucontribute{P.C.L., A.B., and T.S.H. designed the experiments and analyzed the results. GB analyzed the results and provided feedback on the manuscript. P.C.L. composed the quantum programs and executed them in software and on hardware.  H.X. performed the finite size analysis and helped with comparison of results with classical scenarios with the help from B.K.  All authors contributed to writing and approved the presentation of these results.}
\competing{The authors declare that they have no competing interests.}
\funding{This material is based upon work supported by the U.S. Department of Energy, Office of Science, National Quantum Information Science Research Centers, Quantum Science Center.}
\ack{The authors thank Paul Kairys for interesting discussions.  This research used resources of the Oak Ridge Leadership Computing Facility at the Oak Ridge National Laboratory, which is supported by the Office of Science of the U.S. Department of Energy under Contract No.~DE-AC05-00OR22725. P.C.L. and T.S.H. acknowledge support from the Office of Science, Early Career Research Program. A.B., G.B. and B.K. acknowledge funding from the U.S. Department of Energy, Office of Science, National Quantum Information Science Research Centers, Quantum Science Center.}
\bibliographystyle{unsrt}
\bibliography{references}
\end{document}